\documentclass[11pt]{article}

\pdfoutput=1

\usepackage[letterpaper,
            textwidth=6.5in,
            textheight=9.0in,
            marginparsep=0.05in,
            marginparwidth=0.75in,
            heightrounded]{geometry}

\usepackage[T1]{fontenc}
\usepackage[utf8]{inputenc}

\usepackage{mathptmx}          
\usepackage[scaled=0.92]{helvet} 
\usepackage{courier}

\IfFileExists{microtype.sty}{%
  \usepackage[expansion=false,protrusion=true]{microtype}%
}{}

\usepackage{amsmath,amssymb,amsthm}

\usepackage{booktabs}
\usepackage{array}
\usepackage{tabularx}
\usepackage{makecell}
\renewcommand{\arraystretch}{1.15}

\usepackage{graphicx}
\usepackage{float}
\usepackage[font=small,labelfont=bf,labelsep=period,
            skip=6pt]{caption}

\usepackage[dvipsnames]{xcolor}

\usepackage{tikz}
\usetikzlibrary{positioning,arrows.meta,calc,shapes.geometric,backgrounds,fit}

\definecolor{rnfblue}{HTML}{2C5282}
\definecolor{rnfinput}{HTML}{EBF4FA}
\definecolor{rnfproc}{HTML}{FEF5E7}
\definecolor{rnfoutput}{HTML}{E8F5E9}
\definecolor{rnfamber}{HTML}{D68910}
\definecolor{rnfgray}{HTML}{4A5568}
\definecolor{rnfborder}{HTML}{94A3B8}
\definecolor{rnfabs}{HTML}{FDF2F8}
\definecolor{rnfabsborder}{HTML}{BE185D}

\usepackage{titlesec}
\titleformat{\section}
  {\normalfont\Large\bfseries\sffamily}{\thesection}{1em}{}
\titleformat{\subsection}
  {\normalfont\large\bfseries\sffamily}{\thesubsection}{1em}{}
\titleformat{\subsubsection}
  {\normalfont\normalsize\bfseries\sffamily}{\thesubsubsection}{1em}{}
\titlespacing*{\section}{0pt}{14pt plus 4pt minus 2pt}{6pt plus 2pt minus 2pt}
\titlespacing*{\subsection}{0pt}{12pt plus 3pt minus 2pt}{4pt plus 2pt minus 2pt}
\titlespacing*{\subsubsection}{0pt}{10pt plus 2pt minus 2pt}{3pt plus 2pt minus 2pt}

\usepackage[authoryear,round]{natbib}
\setcitestyle{aysep={,}}

\usepackage[colorlinks=true,
            linkcolor=MidnightBlue,
            citecolor=MidnightBlue,
            urlcolor=MidnightBlue,
            breaklinks=true]{hyperref}
\usepackage{url}

\usepackage{enumitem}
\setlist{nosep,leftmargin=2em}

\IfFileExists{orcidlink.sty}{\usepackage{orcidlink}}{%
  \newcommand{\orcidlink}[1]{}
}

\usepackage{fancyhdr}
\fancypagestyle{plain}{%
  \fancyhf{}%
  \fancyfoot[C]{\footnotesize\thepage}%
}

\makeatletter
\renewcommand{\maketitle}{%
  \begingroup
  \centering
  {\LARGE\bfseries\sffamily \@title \par}%
  \vskip 1.5em%
  {\@author \par}%
  \endgroup
  \vskip 1em
  \hrule height 0.4pt
  \vskip 1.2em
}
\makeatother

\title{The Reliance Negotiation Framework:\\[0.2em]
\large A Dynamic Process Model of Student LLM Engagement in Academic Writing}

\author{%
  \textbf{Shahin Hossain}\,\orcidlink{0000-0002-3461-1147}\\
  Department of Language, Literacy and Culture\\
  University of Maryland, Baltimore County\\
  Baltimore, MD 21250, USA\\
  \texttt{shahinh1@umbc.edu}
}
\date{}

\begin{document}
\maketitle
\thispagestyle{plain}

\begin{abstract}
\noindent Student engagement with large language models (LLMs) in academic writing is not a stable trait, an adoption decision, or a competency level; it is a continuously negotiated process that existing frameworks cannot adequately theorize. Typological models provide categories without mechanisms; technology acceptance models explain adoption but not post-adoption quality; AI literacy frameworks treat competency as a static predictor rather than a live input. None accounts for within-student variability across tasks, the developmental paradox whereby experience produces habituation rather than sophistication, or principled non-use as a form of ethical reasoning. This article introduces the Reliance Negotiation Framework (RNF), developed from a sequential explanatory mixed-methods study of 382 undergraduates at a public minority-serving institution in the United States (survey, $N = 382$; 14 semi-structured interviews; three qualitative survey strands; 1{,}435 coded instances). The RNF reconceptualizes LLM reliance as an ongoing negotiation among four concurrent inputs (perceived benefits, perceived risks, ethical commitments, and situational demands) with outputs that recursively modify subsequent decisions. A Two-Model Architecture accommodates the 13.0\% of participants whose categorical ethical commitments foreclose negotiation entirely. The framework generates four falsifiable predictions with implications for AI literacy pedagogy, academic integrity policy, and equity-centered practice at minority-serving institutions.

\vspace{0.6em}
\noindent\textbf{\sffamily Keywords:}\; Reliance Negotiation Framework\; $\cdot$\; LLM reliance\; $\cdot$\; dynamic process model\; $\cdot$\; negotiation-mode\; $\cdot$\; principled non-use\; $\cdot$\; academic writing\; $\cdot$\; minority-serving institution
\end{abstract}

\vskip 0.4em
\hrule height 0.4pt
\vskip 0.8em

\section{Introduction}

\subsection*{Why a New Framework Is Necessary}

Generative AI has entered higher education faster than institutions have developed the conceptual tools to respond to it. Within two years of ChatGPT's public release, universities worldwide had issued policies governing student use of large language models (LLMs) in academic writing, premised on the assumption that such use is a discrete, detectable event that can be permitted or prohibited. That assumption is wrong. Student LLM engagement in academic writing is not a binary, not a stable disposition, and not a literacy level. It is a continuously negotiated process: context-sensitive, developmentally evolving, and shaped by competing beliefs, values, and situational pressures that no current framework adequately models.

The field has generated substantial documentation of LLM adoption patterns \citep{Strzelecki2024}, perceived benefits and cognitive risks \citep{BaidooAnsah2023}, ethical concerns \citep{Eaton2023}, and differential engagement quality \citep{AllenKendeou2024}. What is missing is the theoretical architecture capable of explaining that evidence, answering why students engage as they do, how that changes, and what it means when a student refuses to engage at all. That gap is a structural condition: it limits the field's capacity to design effective interventions, formulate coherent policy, and extend equity-conscious practice to students whose AI engagement carries the highest developmental stakes.

Three theoretical frameworks currently govern the field's explanatory architecture, and each confronts a phenomenon it cannot account for.

\subsection*{Within-Student Variability}

Typological models, including the four-type taxonomy of Strategic, Instrumental, Dialogic, and Dependent reliance derived from the present study, provide categorical descriptions of engagement forms but cannot account for context-sensitive variation within the same student. All 14 interview participants exhibited overlapping reliance orientations across tasks: strategic restraint in one course coexisted with instrumental task-direction in another; dialogic brainstorming in humanities essays coexisted with minimal use in STEM lab reports. Typological models identify the dominant orientation; they cannot explain why the same student exhibits a different one tomorrow.

\subsection*{The Developmental Paradox}

Technology acceptance models, including TAM and UTAUT2, explain adoption but not post-adoption quality \citep{Strzelecki2024}. Greater prior LLM exposure predicts higher reliance intensity ($\beta = .308$) but also, counter-intuitively, membership in all non-Strategic reliance types in the multinomial model (ORs: 2.89--6.83). Experience does not reliably produce sophistication; in many students, it normalizes broader, less regulated use.

\subsection*{Principled Abstention}

In the full open-response qualitative strand ($n = 361$), 13.0\% of respondents ($n = 47$) described a deliberate, values-grounded refusal to use generative AI for academic writing, not from low literacy or low exposure, but from categorical ethical commitment. Typological models collapse these students with low-use low-literacy users; adoption models predict that high utility-value beliefs should produce engagement; AI literacy frameworks \citep{LongMagerko2020,AllenKendeou2024} treat principled refusal as a failure of engagement rather than a form of it.

The Reliance Negotiation Framework (RNF) addresses all three inadequacies. It reconceptualizes LLM reliance as a continuously negotiated process driven by four concurrent inputs (perceived benefits, perceived risks, ethical commitments, and situational demands) whose outputs recursively modify the inputs governing subsequent decisions. This recursive structure produces the developmental trajectories adoption models cannot explain; the situational sensitivity of the negotiation produces the within-student variability typological models cannot account for; and a Two-Model Architecture distinguishes negotiation-mode students from those whose categorical ethical commitments foreclose the negotiation entirely, making principled abstention theorizable rather than invisible.

The RNF advances three claims, each with direct implications for research design, institutional policy, and equity practice:

\begin{description}[leftmargin=2.5em,style=nextline]
  \item[\textbf{C1}] Observable reliance behaviors are the outputs of an ongoing negotiation, not expressions of a stable type, which means that interventions targeting types rather than negotiation inputs will be systematically misdirected.
  \item[\textbf{C2}] The negotiation is recursive: each reliance output modifies the inputs governing subsequent decisions, producing developmental trajectories that are neither inevitable nor linear, which means that the same intervention will produce different outcomes depending on where in the trajectory it is applied.
  \item[\textbf{C3}] A meaningful proportion of students, 13.0\% in this study, operate outside negotiation-mode logic entirely, governed by categorical ethical constraints rather than contextual trade-offs, which means that any measurement instrument or policy architecture built on engagement assumptions will structurally misclassify them.
\end{description}

The remainder of this article proceeds as follows. The theoretical background situates the RNF within three parent frameworks: Expectancy--Value Theory \citep{WigfieldEccles2000}, situative cognition \citep{Greeno1998}, and AI literacy \citep{LongMagerko2020}; and it specifies the explanatory gap each leaves that the RNF addresses. The theoretical architecture develops the four-input negotiation model, the Two-Model Architecture, and the three defining properties of the framework. Two participant trajectories then demonstrate the framework in action. Four falsifiable predictions constitute the RNF's empirical research agenda. The discussion develops implications for AI literacy pedagogy, academic integrity policy, and equity-centered practice at minority-serving institutions.

\section{Theoretical Background and Framework Positioning}

The RNF is built at the intersection of three established research programs (motivational psychology, situated cognition, and AI literacy), each of which reaches a specific explanatory limit when confronted with student LLM use in academic writing. This section specifies what each tradition correctly identifies, where each runs out, and how the RNF's architecture is shaped by both contributions. Table~\ref{tab:frameworks} provides a compact orientation.

\begin{table}[H]
\centering
\caption{Theoretical frameworks integrated in the RNF.}
\label{tab:frameworks}
\small
\renewcommand{\arraystretch}{1.25}
\begin{tabularx}{\textwidth}{@{}p{3.1cm} p{2.6cm} X X@{}}
\toprule
\textbf{Framework} & \textbf{Tradition} & \textbf{Correctly identifies} & \textbf{RNF extension} \\
\midrule
Expectancy--Value Theory \citep{WigfieldEccles2000}
& Motivational psychology
& Motivational beliefs (value, cost, expectancy) as strongest predictors of engagement
& Supplies \emph{perceived benefits} and \emph{perceived risks} inputs; adds recursive feedback \\
\addlinespace
Situative Cognition \citep{Greeno1998}
& Cognitive / ecological psychology
& Context-sensitive learning; affordances and constraints explain within-agent variation
& Supplies \emph{situational demands} input; motivates Two-Model Architecture \\
\addlinespace
AI Literacy Framework \citep{LongMagerko2020}
& HCI / learning sciences
& Competency mediates AI engagement and AI-mediated outcomes
& Competency as dynamic negotiation input, not static antecedent \\
\addlinespace
3P Model \citep{Biggs1993}
& Educational psychology
& Presage--Process--Product as joint determinants of learning outcomes
& Extended from linear to recursive loop: product modifies presage \\
\bottomrule
\end{tabularx}

\vspace{0.5em}
\footnotesize\emph{Note.} Each framework's structural limitation is developed in the prose of Section~2. RNF = Reliance Negotiation Framework.
\end{table}

\subsection{Expectancy--Value Theory: The Motivational Substrate}

The expectancy--value tradition rests on a foundational claim: individuals' choices, persistence, and performance on achievement tasks are jointly determined by their beliefs about how well they will perform and by the subjective value they attach to the task \citep{Atkinson1957,Eccles1983}. Wigfield and Eccles's \citeyearpar{WigfieldEccles2000} elaboration identifies four components of subjective task value (attainment value, intrinsic value, utility value, and cost), together forming a motivational calculus that predicts not only whether students engage but how they allocate effort across competing demands. Longitudinal evidence confirms that these beliefs are not fixed: they shift systematically with evaluative feedback, social comparison, and domain experiences across the school years \citep{Eccles1993,Wigfield1997}.

\subsubsection{What EVT Correctly Identifies}

EVT's core contribution is its cost architecture. Treating cost as a multi-dimensional construct, distinguishing effort cost, opportunity cost, and emotional cost, provides the conceptual vocabulary for understanding why students who perceive high utility value in LLM use do not uniformly engage at high intensity. The costs of engagement (academic integrity risk, cognitive effort of verification, social cost of disclosure) operate as genuine motivational deterrents that vary by task, institutional context, and individual values orientation. The quantitative evidence from the parent study \citep{Hossain2026} confirms EVT's centrality: motivational beliefs were the strongest predictor block in the hierarchical regression ($\beta = .630$, $\Delta R^2 = .256$, $R^2 = .722$), providing strong empirical warrant for treating EVT's constructs as the motivational substrate of LLM reliance decisions.

\subsubsection{Structural Limitation and RNF Extension}

EVT's explanatory power is bounded by two structural assumptions. First, it treats value and cost as independent dimensions in a static calculus. In LLM-assisted writing this independence fails: AI-mediated compression of the drafting process simultaneously produces time savings (utility value) and skill atrophy risk (effort cost, developmental cost). The efficiency--atrophy paradox that recurs in the qualitative data cannot be represented as a simple trade-off between independent dimensions; it is a dialectical relationship in which the magnitude of one is constituted by the magnitude of the other. Second, EVT models the motivational antecedents of a decision at a point in time but provides no recursive architecture through which the outcomes of prior decisions modify subsequent assessments. A semester's experience of AI-assisted writing recalibrates utility value beliefs, cost perceptions, and self-efficacy assessments in ways that shape the next semester's decisions. The RNF maps EVT's value components onto the \emph{perceived benefits} input and EVT's cost dimension onto the \emph{perceived risks} input (disaggregated into four sub-types), while embedding both within a recursive feedback architecture through which negotiation outputs modify the inputs governing subsequent decisions.

\subsection{Situative Cognition: The Contextual Substrate}

The situative perspective, developed most comprehensively by \citet{Greeno1998} and drawing on Lave and Wenger's \citeyearpar{LaveWenger1991} communities of practice framework, Gibson's \citeyearpar{Gibson1986} ecological psychology, and Barwise and Perry's \citeyearpar{BarwisePerry1983} situation theory, shifts the unit of analysis from individual cognitive processes to the activity systems in which agents participate. Knowing is not a property of a mind but a property of a person-in-situation: what individuals know and how they deploy that knowledge is constituted by the affordances and constraints of their activity system \citep[p.~5]{Greeno1998}.

\subsubsection{What Situative Cognition Correctly Identifies}

Three contributions are indispensable to the RNF. First, it explains within-student behavioral variability without invoking inconsistency: a student who exercises strategic restraint in a humanities seminar and instrumental task-direction in a STEM lab is responding to genuinely different affordance--constraint configurations, not behaving erratically. Second, it provides a non-individualistic account of motivation: engagement is a property of an individual's relationship to the practices and identity structures of their communities \citep[p.~9]{Greeno1998}, which explains why norm-following and identity-protecting behaviors both shape reliance decisions. Third, and most importantly for this framework, it treats material and technological systems as constitutive rather than contextual: LLMs do not merely facilitate writing tasks but reshape what counts as a first draft, what counts as authorship, and what counts as adequate research, in ways the activity systems of academic writing have not yet stabilized.

\subsubsection{Structural Limitation and RNF Extension}

The situative framework has no motivational architecture: it can explain why different contexts produce different reliance behaviors but cannot explain why two students in identical situational contexts arrive at qualitatively different decisions. The motivational differentiation EVT explains (differences in utility value beliefs, cost perceptions, and self-efficacy) is outside its scope. More fundamentally, the framework has no category for categorical behavioral refusal. The framework is premised on participation: knowing, learning, and identity are all constituted through participation in activity systems \citep{LaveWenger1991}. Non-participation is theorizable only as peripheral participation or identity contrast, not as a categorical ethical override. The RNF derives the \emph{situational demands} input and the situated property of the negotiation directly from Greeno's affordances-and-constraints architecture, while the Two-Model Architecture addresses the categorical refusal the situative framework cannot theorize.

\subsection{The AI Literacy Framework: The Competence Substrate}

AI literacy, the set of competencies enabling individuals to critically evaluate, collaborate with, and responsibly deploy AI, was formally theorized by \citet{LongMagerko2020} in a framework synthesizing 150 documents across AI education and related literacies. Their 17 core competencies address what AI is, what it can do, how it works, how it should be used, and how people perceive it. \citet{AllenKendeou2024} extended AI literacy theorization to educational contexts, proposing the ED-AI Lit framework, organized around Knowledge, Evaluation, Collaboration, Contextualization, Autonomy, and Ethics, which emphasizes that students must develop the capacity to evaluate AI outputs critically, collaborate effectively with AI systems, and exercise autonomous, ethically grounded judgment in their interactions with AI.

\subsubsection{What AI Literacy Correctly Identifies}

The AI literacy framework's core contribution is identifying competency as the mediating variable between AI engagement and AI-mediated learning outcomes: an insight both EVT and situative cognition miss. EVT explains why students choose to engage; situative cognition explains how context shapes that engagement. Neither explains why two students with identical motivational profiles and situational contexts produce qualitatively different reliance behaviors. The AI literacy framework explains this: the student who verifies has the competencies to evaluate AI outputs against independent evidence; the student who does not lacks them or lacks confidence deploying them. The moderation finding that AI literacy buffers self-efficacy inflation in dependent users ($\beta = -.085$, $p = .027$) provides direct quantitative evidence for this mechanism, confirming that competency recalibrates the perceived benefit input for students who had been overweighting it.

\subsubsection{Structural Limitation and RNF Extension}

The framework's most significant limitation is treating competency as a static predictor rather than a dynamic negotiation input. \citet{LongMagerko2020} position AI literacy as an antecedent condition: more competency produces more productive engagement. This cannot explain why the same student applies systematic verification in high-stakes research writing but omits it in low-stakes posts; competency in the AI literacy framework is a possession the student carries into every context; the RNF treats it as a dimension of the negotiation activated or suppressed by situational demands and motivational inputs. A second limitation: the framework has no account of principled non-use as sophisticated competency deployment. Respondent R141, who articulated refusal grounded in training data provenance, environmental impact, and epistemic insufficiency, was not exhibiting low AI literacy but a form the framework has no category for. Tyler's adversarial verification practice, treating AI outputs as hypotheses to falsify, developed through trial and error rather than formal instruction, represents what the RNF terms \emph{adaptive AI literacy}: competency assembled from accumulated encounters with unreliable outputs, a developmental pathway that is slower, more error-prone, and less generalizable than formal instruction, and inequitably distributed in MSI contexts where formal AI literacy curricula are unevenly available. The RNF incorporates AI literacy as the competency dimension of the negotiation and introduces adaptive AI literacy as a theoretically necessary extension.

\subsection{The Integration Argument}

Each of the three frameworks correctly identifies one dimension of student LLM reliance while being structurally unable to account for the other two. EVT identifies the motivational inputs but has no contextual architecture and no recursive feedback mechanism. Situative cognition identifies the contextual architecture but has no motivational differentiation and no account of categorical refusal. AI literacy identifies the competency mediator but treats it as static and has no account of its developmental acquisition outside formal instruction.

The RNF integrates them by mapping each framework's correct insight onto a distinct dimension of a single negotiation process: EVT supplies the motivational substrate (perceived benefits and perceived risks); situative cognition supplies the contextual substrate (situational demands and the situated property of the negotiation); AI literacy supplies the competency substrate (the accuracy of benefit and risk assessments). Biggs's \citeyearpar{Biggs1993} 3P Model supplies the structural scaffolding, providing the Presage--Process--Product sequence onto which the RNF's recursive feedback loop is mapped, extended from a linear sequence to a recursive loop in which product outcomes modify presage conditions governing subsequent negotiations. What the RNF adds beyond any integration of these three frameworks is a process model: a specification of the mechanism through which motivational, contextual, and competency inputs combine to produce specific reliance outputs at specific moments, and through which those outputs feed back into the inputs governing subsequent decisions. None of the three parent frameworks provides this mechanism, because each was developed to explain a different class of phenomenon.

\section{Theoretical Architecture of the RNF}

\begin{figure}[H]
\centering
\resizebox{\textwidth}{!}{%
\begin{tikzpicture}[
  font=\sffamily\footnotesize,
  node distance=0.35cm and 1.4cm,
  >={Stealth[length=2mm,width=1.8mm]},
  input/.style={
    draw=rnfborder, thin, fill=rnfinput,
    rounded corners=2pt,
    text width=4.2cm, minimum height=1.4cm,
    align=left, inner sep=5pt,
    font=\sffamily\scriptsize
  },
  proc/.style={
    draw=rnfborder, thin, fill=rnfproc,
    rounded corners=2pt,
    text width=4.0cm, minimum height=1.3cm,
    align=left, inner sep=5pt,
    font=\sffamily\scriptsize
  },
  output/.style={
    draw=rnfborder, thin, fill=rnfoutput,
    rounded corners=2pt,
    text width=4.2cm, minimum height=1.3cm,
    align=left, inner sep=5pt,
    font=\sffamily\scriptsize
  },
  core/.style={
    draw=rnfamber, very thick, fill=white,
    rounded corners=3pt,
    text width=4.0cm, minimum height=1.7cm,
    align=center, inner sep=6pt,
    font=\sffamily\small\bfseries, text=rnfamber
  },
  absbox/.style={
    draw=rnfabsborder, thin, fill=rnfabs,
    rounded corners=2pt,
    text width=5.0cm, minimum height=1.1cm,
    align=center, inner sep=5pt,
    font=\sffamily\scriptsize
  },
  colhead/.style={text=rnfblue, font=\sffamily\small\bfseries, align=center},
  arr/.style={-{Stealth[length=2mm]}, draw=rnfgray, thin},
  recarr/.style={-{Stealth[length=2.5mm]}, draw=rnfamber, very thick, dashed}
]

\node[colhead, font=\sffamily\large\bfseries] (title) at (0, 11.2)
  {The Reliance Negotiation Framework (RNF)};

\node[colhead] (hIN)  at (-7.2, 7.1) {INPUTS};
\node[colhead] (hPR)  at ( 0.0, 7.1) {NEGOTIATION PROCESS};
\node[colhead] (hOUT) at ( 7.2, 7.1) {OUTPUTS};
\draw[rnfblue, thin] (-10.5, 6.75) -- (10.5, 6.75);

\node[input] (I1) at (-7.2, 5.1) {%
  {\color{rnfblue}\bfseries Perceived benefits}\\[2pt]
  Time savings \textperiodcentered\ Idea generation\\
  Quality improvement \textperiodcentered\ Scaffolding\\[1pt]
  {\itshape\color{rnfgray} EVT: utility, attainment, intrinsic value}
};
\node[input] (I2) at (-7.2, 2.6) {%
  {\color{rnfblue}\bfseries Perceived risks}\\[2pt]
  Academic integrity \textperiodcentered\ Epistemic\\
  Cognitive-developmental \textperiodcentered\ Social\\[1pt]
  {\itshape\color{rnfgray} EVT: cost dimension (4 sub-types)}
};
\node[input] (I3) at (-7.2, 0.1) {%
  {\color{rnfblue}\bfseries Ethical commitments}\\[2pt]
  Tier 1: compliance-based\\
  Tier 2: norm-dependent\\
  Tier 3: principled {\itshape\color{rnfgray}(Kohlberg, 1981)}
};
\node[input] (I4) at (-7.2, -2.4) {%
  {\color{rnfblue}\bfseries Situational demands}\\[2pt]
  Time pressure \textperiodcentered\ Task stakes\\
  Instructor policy \textperiodcentered\ Disciplinary norms\\[1pt]
  {\itshape\color{rnfgray} Greeno (1998): affordance-constraint}
};

\node[core] (core) at (0, 4.3) {%
  Reliance negotiation\\[3pt]
  {\normalfont\sffamily\scriptsize\color{rnfgray}
  Concurrent \textperiodcentered\ Dynamic \textperiodcentered\ Recursive}
};
\node[proc] (P1) at (0, 1.9) {%
  {\color{rnfamber}\bfseries Recursive}\\[1pt]
  Each output modifies inputs for the next cycle.\\
  {\itshape\color{rnfgray} Biggs (1993): Presage $\to$ Process $\to$ Product}
};
\node[proc] (P2) at (0, 0.0) {%
  {\color{rnfamber}\bfseries Situated}\\[1pt]
  Same student negotiates differently by task / context.\\
  {\itshape\color{rnfgray} Greeno (1998): situative cognition}
};
\node[proc] (P3) at (0, -1.9) {%
  {\color{rnfamber}\bfseries Developmental}\\[1pt]
  Compliance $\to$ principled reasoning over time.\\
  {\itshape\color{rnfgray} Kohlberg (1981): moral development}
};

\node[output] (O1) at (7.2, 5.1) {%
  {\color{rnfblue}\bfseries Reliance behavior}\\[2pt]
  Strategic \textperiodcentered\ Instrumental\\
  Dialogic \textperiodcentered\ Dependent\\[1pt]
  {\itshape\color{rnfgray} Primary output of negotiation}
};
\node[output] (O2) at (7.2, 2.6) {%
  {\color{rnfblue}\bfseries Writing outcomes}\\[2pt]
  Self-efficacy \textperiodcentered\ Quality\\
  Originality \textperiodcentered\ Critical thinking
};
\node[output] (O3) at (7.2, 0.1) {%
  {\color{rnfblue}\bfseries Ethical trajectory}\\[2pt]
  Tier advancement over time.\\[1pt]
  {\itshape\color{rnfgray} Kohlberg (1981) \textperiodcentered\ Long-term}
};

\draw[arr] (I1.east) -- (core.west);
\draw[arr] (I2.east) -- (core.west);
\draw[arr] (I3.east) -- (core.west);
\draw[arr] (I4.east) -- (core.west);
\draw[arr] (core.east) -- (O1.west);
\draw[arr] (core.east) -- (O2.west);
\draw[arr] (core.east) -- (O3.west);

\draw[recarr]
  (O1.north)
  to[out=90, in=90, looseness=0.9]
  node[pos=0.5, above=2pt, font=\sffamily\scriptsize\itshape\bfseries, text=rnfamber]
    {Recursive feedback loop \textperiodcentered\ outputs modify inputs for the next cycle}
  (I1.north);

\node[font=\sffamily\scriptsize\itshape, text=rnfgray, anchor=east]
  at (-1.0, -4.3) {Negotiation-mode: 87.0\% ($n = 314$)};
\node[font=\sffamily\scriptsize\itshape, text=rnfabsborder, anchor=west]
  at ( 1.0, -4.3) {Abstention-mode: 13.0\% ($n = 47$)};
\node[colhead, text=rnfabsborder, font=\sffamily\scriptsize\bfseries]
  at (0, -4.85) {TWO-MODEL ARCHITECTURE \textperiodcentered\ ABSTENTION-MODE BOUNDARY};
\draw[dashed, thick, rnfabsborder] (-10.5, -5.2) -- (10.5, -5.2);

\node[absbox] (AB1) at (-7.2, -6.25) {%
  {\color{rnfabsborder}\bfseries Categorical ethical constraint}\\
  Non-negotiable principle overrides\\ all inputs regardless of context
};
\node[absbox] (AB2) at ( 0.0, -6.25) {%
  {\color{rnfabsborder}\bfseries Negotiation foreclosed}\\
  Lexical constraint dominates;\\ no trade-off occurs
};
\node[absbox] (AB3) at ( 7.2, -6.25) {%
  {\color{rnfabsborder}\bfseries Principled non-use}\\
  Stable refusal independent of\\ situational affordances
};
\end{tikzpicture}
}
\caption{The Reliance Negotiation Framework: four concurrent inputs (left) feed the negotiation process (center), producing three output dimensions (right). The amber dashed arc represents the recursive feedback loop through which each output modifies input weights governing subsequent negotiation cycles. The dashed horizontal boundary separates negotiation-mode students (87.0\%, $n = 314/361$) from abstention-mode students (13.0\%, $n = 47/361$) whose categorical ethical commitments foreclose the negotiation entirely.}
\label{fig:rnf}
\end{figure}
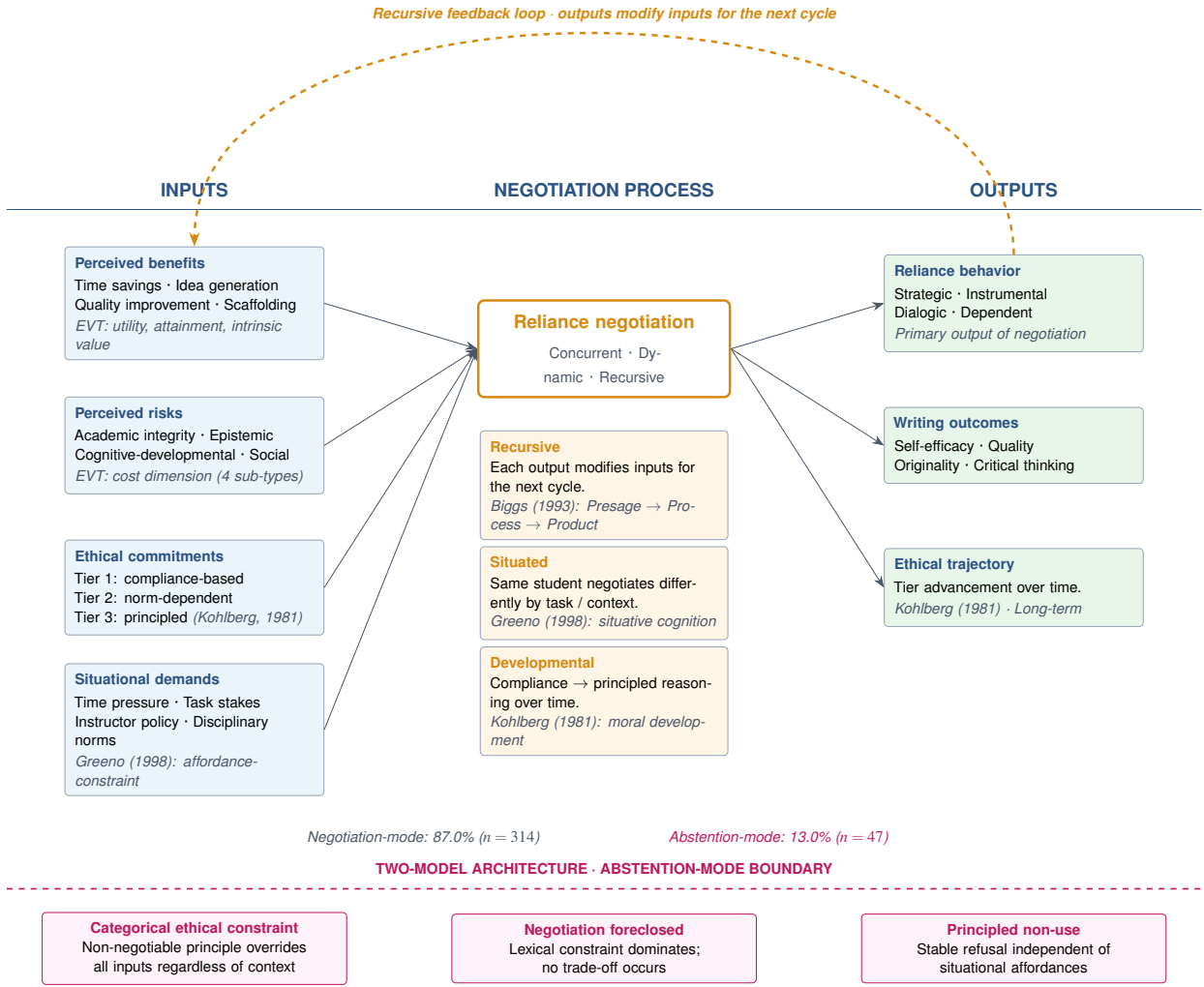

The RNF rests on a single foundational claim: observable LLM reliance behaviors are not expressions of stable individual traits but outputs of an ongoing negotiation process. This section develops that architecture through four moves: the four concurrent inputs and their empirical grounding (Table~\ref{tab:inputs}), the concurrent negotiation mechanism, the Two-Model Architecture distinguishing negotiation-mode from abstention-mode students (Table~\ref{tab:twomodel}), and the three defining properties (Table~\ref{tab:properties}). Figure~\ref{fig:rnf} provides a visual representation of the complete architecture.

\subsection{Core Claim: Reliance as Negotiated Process}

The RNF's foundational claim is that observable LLM reliance behaviors (using AI for brainstorming, submitting AI-assisted text, deploying one LLM to ``humanize'' the output of another, or refusing AI engagement entirely on principle) are the outputs of an ongoing negotiation among four concurrent inputs. Observable behavior is not the phenomenon to explain; it is the outcome whose production the framework is designed to theorize. The inputs, their interaction, and the recursive modification of inputs by prior outputs constitute the explanatory machinery of the framework. Table~\ref{tab:inputs} provides a structured overview of the four inputs before each is developed in the prose below.

\begin{table}[H]
\centering
\caption{The four RNF inputs: definitions, sub-dimensions, parent frameworks, and empirical grounding.}
\label{tab:inputs}
\small
\renewcommand{\arraystretch}{1.25}
\begin{tabularx}{\textwidth}{@{}p{2.6cm} X X p{3.3cm}@{}}
\toprule
\textbf{Input} & \textbf{Sub-dimensions} & \textbf{Parent framework} & \textbf{Empirical evidence} \\
\midrule
Perceived benefits
& Time savings; idea generation; quality improvement; engagement enjoyment
& EVT value components \citep{WigfieldEccles2000}
& $\beta = .630$; $\Delta R^2 = .256$; strongest predictor block \\
\addlinespace
Perceived risks
& Academic integrity; epistemic (hallucination); cognitive-developmental; social
& EVT cost dimension \citep{WigfieldEccles2000}
& 14/14 INT experienced hallucination \\
\addlinespace
Ethical commitments
& Tier 1 compliance; Tier 2 norm-dependent; Tier 3 principled
& \citet{Kohlberg1981}; Three-Tier Ethical Reasoning Model
& Predicts surveillance-sensitivity across qualitative strand \\
\addlinespace
Situational demands
& Time pressure; instructor policy; disciplinary norms; assessment stakes
& \citet{Greeno1998}; \citet{Biggs1993}
& First-gen $\beta = .153$; STEM $\beta = .127$ \\
\bottomrule
\end{tabularx}

\vspace{0.5em}
\footnotesize\emph{Note.} Definitions of each input are developed in the prose below. INT = interview participants; $\beta$ = standardized regression coefficient; EVT = Expectancy--Value Theory.
\end{table}

\subsubsection{Input 1: Perceived Benefits}
Perceived benefits are the expected advantages of LLM engagement: time savings, idea generation, quality improvement, drafting scaffolding, and efficiency gains. This input maps onto EVT's value components: utility value (efficiency dividend), attainment value (performance improvement), intrinsic value (engagement enjoyment); and its primacy is confirmed: motivational beliefs constituted the strongest predictor block in the hierarchical regression ($\beta = .630$, $\Delta R^2 = .256$, $R^2 = .722$). Its weight is modulated by situational demands (time pressure amplifies efficiency benefits) and by perceived risks (hallucination experience attenuates quality improvement benefit), so the same benefit perception produces different reliance outputs across task contexts.

\subsubsection{Input 2: Perceived Risks}
Perceived risks are the anticipated costs of LLM engagement, distributed across four sub-dimensions with distinct temporal structures and situational sensitivities. Academic integrity risk (detection probability, institutional penalty) is immediate and surveillance-amplified. Epistemic risk (hallucination, factual error, credibility compromise) is the most developmentally formative dimension: near-universal hallucination experience across the qualitative strand (14/14 interview participants; 22/35 survey respondents) is the primary mechanism through which students develop verification competencies. Cognitive-developmental risk (skill atrophy, over-reliance) is the most chronically under-weighted sub-dimension because its costs are delayed, invisible in the immediate task context, and actively suppressed by deadline pressure. Social risk (disclosure consequences, peer-norm violation) modulates disclosure willingness. A framework conflating these into a single cost dimension cannot explain why the same student manages academic integrity risk appropriately while systematically underweighting developmental risk.

\subsubsection{Input 3: Ethical Commitments}
Ethical commitments are the normative beliefs governing whether and how LLM engagement is appropriate. This input is structurally tiered: Tier 1 compliance-based reasoning weights detection probability (surveillance-contingent hesitation); Tier 2 norm-dependent reasoning weights instructor policy signals (norm-referential behavior); Tier 3 principled reasoning weights self-authored values stable across surveillance contexts. For abstention-mode students, ethical commitments function not as one of four live inputs but as a lexical constraint that forecloses the negotiation before it begins; this is the architectural distinction necessitating the Two-Model Architecture.

\subsubsection{Input 4: Situational Demands}
Situational demands are the task-specific, temporal, and institutional factors modulating the salience and relative weight of the other three inputs. Deadline proximity elevates efficiency benefits while suppressing skill-atrophy risk salience. Instructor policy clarity shapes compliance-based hesitation while leaving principled reasoning inputs unaffected. Disciplinary norms constitute genuinely different affordance structures: STEM contexts differ qualitatively from humanities contexts in what counts as legitimate AI assistance. First-generation status ($\beta = .153$) and STEM discipline ($\beta = .127$) function as presage-level proxies for what situational demands students experience as salient, consistent with the interpretation that equivalent motivational and competency inputs produce different negotiation outputs in different contexts.

\subsection{The Negotiation Mechanism: Concurrent, Mutual, and Context-Weighted}

The negotiation operates when a student encounters a writing task by simultaneously assessing all four inputs, not sequentially but concurrently and mutually. A tight deadline elevates efficiency benefits while suppressing atrophy risk salience. A strict instructor policy elevates the ethical commitment input while reducing the space available to efficiency temptations. A high-stakes assignment elevates both epistemic risk perception and ethical commitment weight, producing a joint constraint that neither input would produce alone.

The concurrent structure means that identical benefit and risk values can produce different reliance outputs depending on how situational demands configure relative weights. A student with high utility value beliefs and moderate academic integrity risk perception will negotiate toward higher reliance intensity under low surveillance than under explicit monitoring, not because underlying beliefs changed but because the situational demands input reconfigured relative weights. This configurational sensitivity is what typological models cannot represent: a negotiation profile is not a fixed disposition but a context-sensitive weight configuration.

\subsection{The Two-Model Architecture: Negotiation-Mode and Abstention-Mode}

The four-input negotiation model describes negotiation-mode students: those for whom all four inputs interact dynamically to produce context-sensitive reliance outputs. The EVT predictor architecture operates cleanly for this group, applicable to 87.0\% of the full qualitative sample ($n = 314/361$).

The Two-Model Architecture exists because this primary architecture fails to describe 13.0\% of respondents ($n = 47/361$), who described a deliberate, values-grounded refusal that operated not as an extreme case of strategic restraint but as a fundamentally different behavioral architecture in which negotiation did not occur. Respondent R141, who articulated refusal on grounds of training data provenance, environmental harm, and epistemic insufficiency, had applied a non-negotiable principle dominating all contextual considerations regardless of deadline pressure, efficiency benefits, or instructor permissions. Following \citet{Nozick1974}, abstention-mode students treat LLM avoidance as a side constraint: non-tradeable against any benefit regardless of magnitude. Table~\ref{tab:twomodel} specifies the architectural distinction across seven dimensions.

\begin{table}[H]
\centering
\caption{The Two-Model Architecture: negotiation-mode and abstention-mode behavioral logic compared.}
\label{tab:twomodel}
\small
\renewcommand{\arraystretch}{1.25}
\begin{tabularx}{\textwidth}{@{}p{3.2cm} X X@{}}
\toprule
\textbf{Dimension} & \textbf{Negotiation-mode} & \textbf{Abstention-mode} \\
\midrule
Behavioral logic
& Marginal trade-offs among four inputs produce context-sensitive outputs
& Categorical ethical constraint forecloses negotiation before inputs are weighed \\
\addlinespace
Role of ethical commitments
& Active input, weighted against benefits, risks, and situational demands
& Lexical constraint dominating all other inputs unconditionally \\
\addlinespace
Response to situational pressure
& Deadline pressure, instructor permissions, and efficiency gains shift output
& Situational affordances do not alter output; constraint is non-negotiable \\
\addlinespace
EVT applicability
& Applies cleanly: utility value $\to$ intensity; AI literacy $\to$ type quality
& Inapplicable: high utility-value beliefs do not produce engagement \\
\addlinespace
Moral reasoning analog
& Tiers 1--3 of the Three-Tier Model; contextually applied reasoning
& Values-positioning outside the cost--benefit moral calculus \\
\addlinespace
Measurement challenge
& Standard instruments capture this population
& Conflated with low-exposure students; requires dedicated abstention scales \\
\addlinespace
Empirical prevalence
& 87.0\% ($n = 314/361$)
& 13.0\% ($n = 47/361$) \\
\bottomrule
\end{tabularx}

\vspace{0.5em}
\footnotesize\emph{Note.} Prevalence figures from the full-sample qualitative strand. EVT = Expectancy--Value Theory.
\end{table}

The measurement consequence is immediate: current instruments conflate three distinct near-zero-score populations (strategic restrainers, principled non-users, and low-exposure students) who respond to institutional interventions in fundamentally different ways. Future instrument development must include validated abstention scales that assess the grounds for near-zero reliance scores.

\subsection{Three Defining Properties: Recursive, Situated, and Developmental}

Three properties together distinguish the RNF from all static typological models and from existing dynamic frameworks. Table~\ref{tab:properties} provides a compact overview before each property is developed.

\begin{table}[H]
\centering
\caption{The three defining properties of the RNF.}
\label{tab:properties}
\small
\renewcommand{\arraystretch}{1.25}
\begin{tabularx}{\textwidth}{@{}X X X@{}}
\toprule
\textbf{R --- Recursive} & \textbf{S --- Situated} & \textbf{D --- Developmental} \\
\midrule
Each output feeds back into the inputs governing subsequent decisions; product modifies presage.
& Same student negotiates differently across task contexts; different activity systems afford different configurations.
& Negotiation matures over time from reactive Tier 1 toward reflective Tier 3 reasoning; trajectory is neither inevitable nor linear. \\
\addlinespace
\textit{Evidence:} Tariq's 87-page disclosure $\to$ selective disclosure next semester. Destiny's learning-impoverishment $\to$ strategic restraint.
& \textit{Evidence:} Tyler's adversarial verification in engineering writing; all 14 INT showed overlapping orientations across tasks.
& \textit{Evidence:} $\beta = .308$ intensity; ORs 2.89--6.83 for non-strategic types; encounter with AI failure as developmental catalyst. \\
\bottomrule
\end{tabularx}

\vspace{0.5em}
\footnotesize\emph{Note.} $\beta$ = standardized coefficient; ORs = odds ratios from multinomial logistic regression; INT = interview participant.
\end{table}

\subsubsection{The Framework Is Recursive}
Each reliance output feeds back into the inputs governing subsequent negotiations, modifying their weights in ways that persist into future task contexts. The mechanism operates through three channels: benefit recalibration (discovering that AI-assisted drafting produces work one cannot defend recalibrates utility value downward); risk recalibration (hallucination encounter develops epistemic risk perception from abstract awareness into calibrated assessment); and ethical tier advancement (Tariq's 87-page disclosure document catalyzed a shift from Tier 1 compliance toward Tier 2 judgment-governed reasoning). Extending Biggs's \citeyearpar{Biggs1993} Presage--Process--Product sequence from linear to recursive (the product modifies presage), a claim supported by the moderation finding that prior exposure attenuates dependent reliance--self-efficacy associations ($\Delta R^2 = .012$, $\beta = -.135$, $p = .003$).

\subsubsection{The Framework Is Situated}
The negotiation process is constituted by the affordance structure of the activity system in which it occurs, not merely influenced by it \citep{Greeno1998}. Tyler's adversarial verification practice, developed through engineering writing where AI mathematical errors are propositionally detectable, illustrates the situated property directly: what constitutes adequate verification is itself constituted by disciplinary epistemic standards, which differ structurally between STEM and humanities contexts.

\subsubsection{The Framework Is Developmental}
The negotiation matures from reactive, fear-based Tier 1 engagement toward reflective, values-based Tier 3 engagement, but the trajectory is neither inevitable nor linear. Greater prior LLM exposure predicts higher reliance intensity ($\beta = .308$) but also membership in all non-Strategic reliance types (ORs: 2.89--6.83). When engagement is not accompanied by verification encounters, reflective feedback, or structured ethical reasoning, the recursive loop recalibrates toward habituation rather than sophistication. Because reliance behaviors are outputs of a recursive developmental process, interventions targeting the inputs (AI literacy education, structured ethical reasoning curricula, transparent institutional policy) can change the trajectory. This is the foundation on which the discussion section's policy recommendations rest.

\section{The RNF in Action: Two Developmental Trajectories}

This section demonstrates the RNF's architecture through two participant trajectories: Tariq, a History senior whose two-semester account illustrates the recursive feedback mechanism and institutional policy as negotiation input; and Destiny, a Psychology sophomore whose trajectory illustrates the efficiency--atrophy paradox. Comparing the two cases reveals how the same framework architecture produces different developmental outcomes depending on which inputs are most salient and which recalibration channel activates first. Participant accounts are drawn from the qualitative strand of the parent study (14 semi-structured interviews). Names are pseudonyms.

\subsection{Tariq: Institutional Policy as Negotiation Input}

Tariq's account covered two consecutive semesters. During the summer course, the \emph{situational demands} input was dominated by a compressed timeline and an instructor requiring full AI prompt-and-response documentation alongside every assignment. Tariq described seeking ``a heavier framework'' from AI under acute time pressure, elevating the \emph{perceived benefits} input beyond its baseline level: the efficiency dividend of AI was objectively larger under compressed conditions. His \emph{ethical commitment} input operated at the Tier 1/Tier 2 boundary: compliance-detection reasoning plus instructor norm-deference, without self-authored academic values. His \emph{perceived risk} input was active but imprecise: he recognized hallucination risk abstractly but had not developed calibrated verification practices, and cognitive-developmental risk did not enter the negotiation at all. The four inputs produced a specific output: substantial AI use disclosed comprehensively as required, yielding an 87-page prompt-and-response document.

The recursive feedback loop then activated. The 87-page document recalibrated Tariq's disclosure cost assessment upward: comprehensive disclosure was now experienced as disproportionately burdensome. In the fall, a different instructor required only ``just a couple of lines under your paper,'' shifting the situational demands input toward proportionate norms. Tariq's negotiation produced qualitatively different output: selective, proportional, judgment-governed disclosure. The same student, with unchanged EVT beliefs and AI literacy level, produced radically different behaviors because one input had changed. Tariq's trajectory demonstrates the RNF's most practically significant institutional claim: policy does not merely observe student negotiation outcomes; it shapes the input weights that govern them.

\subsection{Destiny: The Efficiency--Atrophy Paradox and Developmental Recalibration}

In her high school years, Destiny's \emph{perceived benefits} input was large and immediately confirmed: AI assistance objectively improved her grades. The negotiation output was rational: continued and expanding AI use, consistent with an accurate calculation of immediate benefits. The paradox was not produced by irrationality. It was produced by a rational response to an accurate immediate calculation that failed to perceive a delayed cost: cognitive-developmental risk was structurally invisible in the immediate negotiation context where the grade signal dominated. Academic integrity risk was salient; epistemic risk was present; but the risk that AI-assisted drafting was substituting for the skill development the assignments were designed to produce did not enter the negotiation at all.

The recursive feedback loop is activated through accumulated encounters with the learning cost. ``I used to, when I was in high school, use AI to do almost all of my assignments, and then eventually I started to truly realize that I was not learning anything, so I stopped.'' This encounter activated the developmental-cost dimension of the perceived risk input. By the time Destiny arrived at the MSI as a sophomore, subsequent negotiations operated under a recalibrated input configuration: benefits input remained recognized, but developmental risk was now live and weighted against it, producing the trajectory from dependent toward strategic reliance that the RNF predicts when this recalibration occurs.

\subsection{Cross-Case Analysis}

Table~\ref{tab:cases} maps both trajectories onto the RNF's input architecture.

\begin{table}[H]
\centering
\caption{RNF input architecture applied to Tariq and Destiny.}
\label{tab:cases}
\small
\renewcommand{\arraystretch}{1.25}
\begin{tabularx}{\textwidth}{@{}p{2.8cm} X X@{}}
\toprule
\textbf{Input} & \textbf{Tariq (Tier 1--2)} & \textbf{Destiny (recalibration)} \\
\midrule
Perceived benefits
& High: ``heavier framework'' under time pressure
& Initially high: grade improvement confirmed \\
\addlinespace
Perceived risks
& Integrity: active/Tier 1. Epistemic: abstract. Developmental: not perceived
& Epistemic: low initially. Developmental: invisible until learning encounter \\
\addlinespace
Ethical commitments
& Tier 1/2: compliance + norm-deference
& Tier 1 $\to$ Tier 2; nascent intellectual-growth concern \\
\addlinespace
Situational demands
& Summer: compressed + full citation. Fall: proportionate citation norm
& Year 1: low stakes. Year 2+: higher-stakes disciplinary writing \\
\addlinespace
Negotiation output
& Summer: heavy use + 87-page disclosure. Fall: selective, proportional disclosure
& Year 1: dependent reliance. Year 2+: strategic restraint \\
\addlinespace
Recursive channel
& 87-page doc $\to$ disclosure cost recalibrated $\to$ selective disclosure
& Learning-impoverishment encounter $\to$ developmental risk activated $\to$ strategic restraint \\
\addlinespace
Developmental trajectory
& Compliance $\to$ proportionality judgment (Tier 1--2)
& Tier 1 $\to$ Tier 2 $\to$ nascent Tier 3 \\
\bottomrule
\end{tabularx}

\vspace{0.5em}
\footnotesize\emph{Note.} Input configurations reconstructed from interview accounts. Tier designations follow the Three-Tier Ethical Reasoning Model.
\end{table}

Three observations follow from the comparison that are not available from either case alone. First, the efficiency--atrophy paradox operates through the same structural mechanism in both cases (cognitive-develop\-mental risk invisible within the immediate negotiation), but activates the recursive loop through different channels: institutional policy encounter for Tariq, learning encounter for Destiny. A single-case analysis would have obscured this. Second, these constitute different intervention targets: changing Tariq's trajectory requires changing the situational demands input through coherent disclosure policy; changing Destiny's trajectory requires either formal developmental risk instruction or assignment designs that accelerate the encounter-based recalibration pathway. Third, both students are unambiguously negotiation-mode throughout their trajectories: even at their most dependent, their behaviors are outputs of a negotiation among live inputs, not expressions of categorical ethical constraints, confirming that an instrument conflating their contextual restraint with abstention-mode principled non-use would structurally misclassify both.

Together the two trajectories demonstrate that the RNF's explanatory machinery runs on real data, producing specific accounts of specific behavioral events and generating intervention implications that alternative frameworks, which would classify Tariq as a ``strategic user'' without explaining the mechanism and Destiny as ``dependent then strategic'' without explaining the transition, cannot provide.

\section{Four Testable Predictions}

The RNF generates four falsifiable predictions that constitute a specific research agenda. Following Meehl's \citeyearpar{Meehl1978} argument that a theoretical framework earns scientific credibility by surviving risky tests rather than accumulating confirmatory evidence, each prediction specifies a formal statement, the study design required to test it, the disconfirmation threshold, and the theoretical revision the framework would require if the prediction fails. Table~\ref{tab:predictions} provides a compact summary; the prose below develops each prediction in full.

\begin{table}[H]
\centering
\caption{Summary of the four RNF predictions.}
\label{tab:predictions}
\small
\renewcommand{\arraystretch}{1.25}
\begin{tabularx}{\textwidth}{@{}c X X X@{}}
\toprule
\textbf{P} & \textbf{Prediction (abbreviated)} & \textbf{Required design} & \textbf{Disconfirmation threshold} \\
\midrule
1 & Within-student ICC $< .50$ across task contexts within same course
  & ESM: $\geq 5$ assignments; $\geq 20$ students
  & ICC $\geq .50$ with $p < .05$ \\
\addlinespace
2 & Ethical tier advancement yields surveillance-independent stability (Cohen's $d \geq 0.40$)
  & Pre--post RCT; integrity curriculum; high/low surveillance
  & $d < 0.20$ post-intervention \\
\addlinespace
3 & AI literacy intervention reduces self-efficacy inflation ($\Delta\beta \geq .05$)
  & Controlled pre--post with reliance type and moderation re-estimate
  & No shift in reliance types; moderation unchanged \\
\addlinespace
4 & MSI $\times$ preparation interaction on efficiency--atrophy tipping point
  & Multi-site ($\geq 4$); longitudinal reliance + outcomes
  & No significant institutional moderation \\
\bottomrule
\end{tabularx}

\vspace{0.5em}
\footnotesize\emph{Note.} The theoretical revision required if each prediction fails is developed in the prose below. ESM = experience-sampling methodology; RCT = randomized controlled trial; ICC = intraclass correlation coefficient; MSI = minority-serving institution.
\end{table}

\subsection{Prediction 1: Within-Student Variability}

Reliance behaviors will exhibit significant within-student variability across task contexts within the same course, with an intraclass correlation coefficient (ICC) below .50 when reliance decisions are clustered by student across assignments. The prediction requires a prospective experience-sampling methodology (ESM) design capturing reliance decisions across a minimum of five assignments in a single course, using process-visible data collection: screen recordings, digital revision logs, or process-visible assessments. The ICC statistic is the appropriate analytic tool because it directly quantifies the proportion of variance attributable to between-student differences (trait variance) versus within-student task-to-task variation (situational variance). The RNF predicts that situational variance dominates, producing ICC $< .50$. The framework would be disconfirmed by ICC $\geq .50$ with $p < .05$ across multiple courses and populations, indicating that reliance is trait-like rather than situationally constituted. If Prediction 1 fails, the RNF must revise its treatment of the situational demands input: the framework would require significant downweighting of that input relative to motivational and competency inputs, moving the architecture closer to a modified EVT account with situational moderators than to a full situative negotiation model.

\subsection{Prediction 2: Ethical Tier Advancement and Surveillance-Independent Stability}

Ethical tier advancement produced by structured academic integrity interventions will reduce the surveillance-sen\-sitivity of reliance behavior, producing Cohen's $d \geq 0.40$ in the difference between high-surveillance and low-surveillance reliance consistency scores, measured pre-to-post intervention. The prediction requires a randomized controlled trial (RCT) in which a structured ethical reasoning curriculum is compared against an active control condition, with reliance behavior measured under explicit-monitoring and independent-work conditions at both time points. The primary outcome is the behavioral consistency gap across surveillance conditions; the prediction is that the treatment group shows a significantly larger reduction in surveillance-sensitivity than controls. The framework would be disconfirmed by $d < 0.20$ post-intervention, especially if measured ethical tier advances while behavioral surveillance-sensitivity is unchanged: a finding that would indicate the ethical commitments input is operationally decoupled from reliance behavior outputs. If Prediction 2 fails, the RNF must distinguish expressed ethical tier (self-assessed) from enacted ethical tier (revealed under reduced surveillance), requiring a two-track revision of the tier architecture consistent with Argyris and Sch\"{o}n's \citeyearpar{ArgyrisSchon1978} espoused-values / values-in-use distinction.

\subsection{Prediction 3: AI Literacy Intervention Effects on Negotiation Output Quality}

AI literacy interventions targeting the accuracy of benefit and risk assessments will produce measurable shifts in negotiation output quality: specifically, a reduction in self-efficacy inflation among dependent users, operationalized as a shift in the moderation coefficient from its baseline ($\beta = -.085$, $p = .027$) of at least $\Delta\beta \geq .05$, and increased verification-accompanied reliance among previously uncritical users. The prediction requires a controlled pre--post design in which a validated AI literacy curriculum is administered and reliance type classification is re-assessed alongside a behavioral verification measure. The moderation analysis from the parent study must be re-estimated post-intervention to assess whether the intervention has shifted the coefficient that identifies AI literacy as recalibrating overweighted benefit inputs. The framework would be disconfirmed by a null finding on both outcomes (no shift in reliance type distribution and a statistically unchanged moderation pattern), which would support the AI literacy framework's own account of competency as a static antecedent rather than the RNF's formulation of it as a live negotiation input. If Prediction 3 fails, AI literacy would be repositioned as a moderator of the benefit and risk inputs rather than as a constituent of them, preserving the negotiation architecture while substantially weakening the framework's interventional implications.

\subsection{Prediction 4: Institutional Moderation of Efficiency--Atrophy Tipping-Point Effects}

Negotiation outcomes will be moderated by institutional type and student preparation background, with students at MSIs showing a steeper efficiency--atrophy tipping-point effect: the positive relationship between reliance intensity and skill-substitution outcomes will be significantly stronger at MSIs than at more selective institutions after controlling for individual-level AI literacy and EVT beliefs. The prediction requires a multi-site comparative design ($\geq 4$ institutions varying on MSI status, selectivity, first-generation percentage, and region) with consistent reliance measurement, writing quality assessment, and longitudinal follow-up. The critical term is the MSI status $\times$ reliance intensity interaction on the skill-substitution outcome, estimated in a multilevel model. The parent study provides directional support: first-generation status ($\beta = .153$) and lower-SES ($\beta = .099$, $p = .040$) both predict higher reliance intensity, consistent with the preparation-gap compensation dynamic. The framework would be disconfirmed by a non-significant MSI $\times$ reliance intensity interaction across multiple institutional comparisons, indicating that the tipping point is equally steep regardless of preparation background. If Prediction 4 fails, the MSI context is demoted from a theoretically necessary architectural condition to an empirically contingent feature of the parent study's population, and the equity argument embedded in the framework (that MSI students face structurally steeper substitution risk) would require a different empirical foundation.

\subsection{The Predictions as a Research Programme}

The four predictions are not independent empirical questions. They constitute an integrated research programme in which each tests a distinct architectural claim: Prediction 1 tests the situatedness claim; Prediction 2 tests the ethical tier architecture; Prediction 3 tests the recalibration mechanism of the AI literacy competency dimension; Prediction 4 tests the MSI equity claim. Confirmation of all four would constitute strong support for the RNF's full architecture; partial confirmation would identify which architectural elements are empirically robust and which require revision; full disconfirmation would establish that LLM reliance is better modeled as a trait-based construct with situational moderators than as a continuously negotiated process: a conclusion that would redirect the field's intervention strategy accordingly.

The longitudinal evidence required to directly test these predictions remains to be generated. The parent study's cross-sectional design provides theoretical warrant for each prediction (within-student variability through retrospective interview accounts, ethical tier through qualitative coding, AI literacy's moderation effect through the quantitative moderation analysis, institutional context through first-generation and SES predictors), but prospective longitudinal tests are necessary before any claim of empirical confirmation can be made. The RNF is offered as a framework with strong theoretical grounding and preliminary empirical support, whose full empirical status awaits the research programme these predictions define.

\section{Discussion: Theoretical Implications and Boundary Conditions}

The RNF's theoretical contributions are most clearly visible not in what it adds to existing frameworks but in what it makes possible that was previously unavailable: a mechanism for explaining within-student variability, a recursive architecture that accounts for developmental trajectories, and a Two-Model Architecture that makes principled abstention theorizable. The Discussion draws out the practical consequences of these contributions across four domains: what the RNF enables for the field, what the MSI context requires of equity-conscious AI engagement research, implications for AI literacy pedagogy, and implications for academic integrity policy.

\subsection{What the RNF Enables}

The RNF's most immediate practical consequence is the transformation of reliance \emph{types} into reliance \emph{inputs}. A student producing dependent reliance outputs because of inflated utility value beliefs requires a different intervention than a student producing identical dependent outputs because of Tier 1 compliance-based ethical reasoning that collapses under low surveillance, yet typological classification alone cannot distinguish these students. The RNF's input architecture can: the research question shifts from ``what type is this student?'' to ``which input is producing non-strategic outputs, and how can it be recalibrated?'' For technology acceptance research, the RNF's recursive feedback loop fills the post-adoption gap: two students beginning a course with equivalent EVT profiles will diverge over time if early reliance outputs produce different recalibrations, and the RNF specifies the mechanism: whether the recursive loop runs through active recalibration or passive normalization depends on prior AI literacy, ethical tier, and institutional feedback quality. For AI literacy research, the RNF resolves an empirical anomaly: higher AI literacy simultaneously predicts strategic type membership \emph{and} higher reliance intensity because more accurate benefit assessment produces more selective engagement, not less engagement. This means AI literacy interventions should be evaluated on reliance quality (verification rates, output calibration) rather than reliance frequency.

\subsection{The MSI Context as Theoretical Condition and Equity Imperative}

The RNF was developed at an MSI where 33.5\% of the study sample identified as first-generation college students, preparation heterogeneity is structural, and AI literacy resources are unevenly distributed. These conditions are not incidental to the framework's architecture: three MSI-specific dynamics bear directly on the four inputs and require different institutional responses than frameworks developed at selective institutions would prescribe.

The \emph{perceived benefits} input is amplified by scarcity. Students compensating for preparation gaps perceive AI's efficiency dividend as objectively larger than students augmenting existing competence, a compensation dynamic evidenced in the data (first-generation $\beta = .153$; lower-SES $\beta = .099$, $p = .040$). Addressing this requires building the foundational writing competencies that would shift the efficiency calculus, not reducing perceived AI value.

The \emph{AI literacy competency} dimension is unevenly distributed by institutional context, not individual ability. Students arriving at MSIs without formal AI literacy preparation develop verification competencies through the encounter-based pathway (accumulated experience of AI failure), which is slower, more error-prone, and less generalizable than formal instruction. MSIs require institutional infrastructure: systematic AI literacy integration across first-year curricula, discipline-specific verification training in writing-intensive courses, and faculty professional development enabling instructors to model calibrated AI use.

The \emph{efficiency--atrophy tipping point} operates asymmetrically by preparation level: students with strong foundational skills can augment without substituting; students compensating for preparation gaps are structurally more likely to cross into substitution because the gap AI is filling is precisely the competency whose development AI substitution prevents. The recursive feedback loop amplifies this asymmetry: early substitutive reliance reduces foundational skill investment, which increases subsequent AI efficiency value, which produces higher substitutive reliance. A student who begins a semester at the substitution threshold may end it well past it, not through any failure of motivation or intelligence, but through the recursive dynamics of a negotiation operating without adequate institutional scaffolding.

These three dynamics constitute an equity imperative: frameworks developed and validated at selective institutions should not be applied to MSI populations without MSI-specific validation. A field that theorizes AI engagement from the experience of the most advantaged students and applies those theories universally is not producing general knowledge; it is producing knowledge that systematically misrepresents the students whose AI engagement carries the highest developmental stakes.

\subsection{Implications for AI Literacy Pedagogy}

AI literacy pedagogy aligned with the RNF targets negotiation accuracy (calibrating students' benefit and risk assessments against actual AI performance in their disciplinary writing contexts) rather than engagement frequency or generic ethical awareness. This requires discipline-specific instruction: a student who has learned that LLMs hallucinate in general has not developed the verification competency needed to detect a hallucinated legal citation in a policy brief. Task-specific calibration exercises (structured comparisons of AI outputs against authoritative sources, with documentation of error rates by task type) are the pedagogical intervention the RNF specifies.

Institutions cannot wait for encounter-based learning to produce verification competency through slow accumulation of AI failures, but they can design assignments that accelerate the encounter-based pathway: exercises requiring students to document and analyze AI errors develop the diagnostic competency that formal instruction alone does not reliably produce. Tyler's adversarial verification practice from the parent study (treating AI outputs as hypotheses to falsify rather than answers to accept) illustrates the target level of adaptive AI literacy the RNF specifies.

The most consequential pedagogical distinction the RNF draws is between literacy-centered and compliance-centered instruction. Compliance-centered instruction targets the ethical commitment input's Tier 1 channel and produces reliance behavior contingent on perceived detection probability; when surveillance decreases, compliance-based restraint dissolves. Literacy-centered instruction develops the benefit and risk assessment accuracy that enables informed negotiation decisions regardless of monitoring context. The pedagogical priority is not suppressing engagement but developing the negotiation competency that makes engagement productive rather than substitutive.

\subsection{Implications for Academic Integrity Policy}

Institutions do not merely observe the outcomes of student reliance negotiations; through surveillance environments, policy signals, and disclosure norms, they actively shape the input weights that govern negotiation outputs. Table~\ref{tab:policy} specifies the mechanisms, predicted effects, and critical limitations of three policy approaches derivable from the RNF's architecture.

\begin{table}[H]
\centering
\caption{Three policy approaches to student LLM reliance.}
\label{tab:policy}
\small
\renewcommand{\arraystretch}{1.25}
\begin{tabularx}{\textwidth}{@{}p{2.8cm} X X X@{}}
\toprule
\textbf{Policy type} & \textbf{RNF mechanism targeted} & \textbf{Predicted effect} & \textbf{Critical limitation} \\
\midrule
Detection-centered
& Tier 1 compliance channel of ethical commitment input
& Reduces reliance in monitored contexts only
& Surveillance-dependent; may increase strategic deception \\
\addlinespace
Literacy-centered
& Accuracy of benefit and risk assessments; competency dimension
& More adaptive outputs; reduced self-efficacy inflation
& Leaves ethical tier unaddressed; requires sustained curriculum \\
\addlinespace
Ethics-reasoning-centered
& Ethical tier advancement; Tier 3 principled reasoning capacity
& Stable behavior absent enforcement; surveillance-independent restraint
& Most demanding to implement; slow-developing; longitudinal assessment needed \\
\bottomrule
\end{tabularx}

\vspace{0.5em}
\footnotesize\emph{Note.} Examples of each policy type and fuller discussion of effects appear in the prose. Most real implementations combine elements of all three.
\end{table}

The three approaches are not alternatives on a spectrum from permissive to restrictive. They target fundamentally different mechanisms and produce fundamentally different behavioral outcomes. A detection-centered policy acts on the Tier 1 compliance channel; the student who reduces AI use to avoid detection has not developed more accurate benefit or risk assessments and reverts to prior outputs when monitoring decreases. A literacy-centered policy improves competency inputs but leaves ethical tier unaddressed. An ethics-reasoning-centered policy is the most durable, producing reliance behaviors stable in the absence of enforcement, and also the most demanding: ethical tier advancement is visible only in reliance behavior patterns across contexts and surveillance conditions over time, not in any single assignment.

For MSI contexts specifically, detection-centered policy is not merely insufficient; it is actively inequitable. Students whose AI engagement is driven by compensation dynamics have more to lose from surveillance-induced restraint: the preparation gap AI is filling does not disappear when AI use is restricted; it becomes a preparation gap without a compensatory tool. MSI institutions that adopt detection-centered policies without simultaneously investing in literacy-centered and ethics-reasoning-centered infrastructure are protecting the appearance of traditional academic norms while leaving the structural conditions that produced both the preparation gap and the compensation dynamic entirely unaddressed.

\subsection{Limitations and Future Development}

Three limitations bound the current framework's empirical status. Table~\ref{tab:limitations} specifies each limitation's nature, what it does not invalidate, and the future development it demands.

\begin{table}[H]
\centering
\caption{Three limitations of the RNF and future development required.}
\label{tab:limitations}
\small
\renewcommand{\arraystretch}{1.25}
\begin{tabularx}{\textwidth}{@{}p{3.2cm} X X@{}}
\toprule
\textbf{Limitation} & \textbf{Nature of constraint} & \textbf{Future development required} \\
\midrule
1. Cross-sectional, single-institution design
& No longitudinal data; within-student variability rests on retrospective accounts
& Longitudinal ESM; multi-site replication; prospective input recalibration measurement \\
\addlinespace
2. Retrospective, single-informant data
& Self-report subject to social desirability bias; no triangulation with process data
& Screen recordings, keystroke logs, revision histories; behavioral verification measures \\
\addlinespace
3. No validated abstention measures
& Existing instruments conflate strategic restrainers, principled non-users, and low-exposure students
& Motivation-structure items distinguishing grounds for near-zero reliance; abstention scale validation \\
\bottomrule
\end{tabularx}

\vspace{0.5em}
\footnotesize\emph{Note.} What each limitation does not invalidate is developed in the prose below. ESM = experience-sampling methodology.
\end{table}

The most theoretically consequential limitation is the absence of validated abstention measures. The Two-Model Architecture's most important practical claim (that abstention-mode students respond to institutional interventions differently from negotiation-mode students and low-exposure students) cannot be tested until these three near-zero-reliance populations can be quantitatively distinguished. The required items are not reliance-frequency items at lower thresholds but motivation-structure items assessing the grounds for near-zero reliance: contextual restraint, categorical ethical commitment, or insufficient exposure. Developing and validating this three-way distinction is the single most important methodological advance the RNF requires before its Two-Model Architecture can be tested at scale. The three limitations together define a research agenda rather than a set of objections: a framework that specifies exactly what longitudinal evidence would confirm its recursive mechanism, what process-visible data would triangulate its retrospective qualitative findings, and what instrument development would make its Two-Model Architecture quantitatively testable is a framework that knows what it does not yet know.

\section{Conclusion}

The RNF has established what the field of AI-in-education has so far lacked: not more evidence of what student LLM reliance looks like, but a theoretical account of why it takes the forms it does. By reconceptualizing reliance as the output of a continuously negotiated process among four concurrent inputs (perceived benefits, perceived risks, ethical commitments, and situational demands), the framework explains three phenomena that static typological, motivational, and competency accounts cannot: why the same student engages differently across task contexts, why experience sometimes produces habituation rather than sophistication, and why 13.0\% of students in this study operate outside the negotiation framework entirely, governed by categorical ethical constraints rather than contextual trade-offs. The framework's recursive architecture specifies not merely what reliance behaviors look like at a point in time but how they develop, and what recalibrates them.

What the framework ultimately reveals about institutional policy is this: the choice among policy approaches is not a choice among levels of restrictiveness. It is a choice about what institutions believe student reliance is. Institutions that treat reliance as a prohibited behavior respond with detection and enforcement, and produce compliance-contingent restraint that dissolves when monitoring decreases. Institutions that treat reliance as an information-processing problem respond with literacy curricula, and produce students who engage with AI more accurately, though not necessarily less often. Institutions that treat reliance as a developmental process in a moral domain respond with ethical reasoning education, and produce students whose reliance decisions are stable in the absence of enforcement, because they are governed by internalized values rather than external surveillance. The third response is the most demanding to implement and the slowest to produce visible behavioral change. It is also the only one whose effects persist beyond the institutional context that produced them.

The era of generative AI in higher education has arrived without waiting for institutions to develop coherent responses. The students who most need those responses (first-generation students negotiating with amplified efficiency benefits and encounter-based literacy pathways, students at minority-serving institutions where the efficiency--atrophy tipping point operates asymmetrically by preparation level) cannot afford to wait for the field to generate the longitudinal evidence that would confirm or revise what this framework has proposed. The RNF does not ask institutions to wait. It asks them to act on the best available theoretical account of how student reliance decisions are made and how they develop, and to build the pedagogical infrastructure, policy environments, and equity-conscious interventions that treat those processes not as threats to be suppressed but as developmental capacities to be cultivated.


\end{document}